\documentclass[twocolumn,showpacs,amsfonts,aps,prc,floatfix,superscriptaddress,nofootinbib,groupedaddress]{revtex4-1}

\usepackage{graphicx}  
\usepackage{dcolumn}   
\usepackage{bm}        
\usepackage{amssymb}   
\usepackage{wasysym}
\usepackage{hyperref}
\hypersetup{colorlinks=true}
\usepackage{graphicx}
\usepackage{epstopdf}
\usepackage{graphicx}
\usepackage{dcolumn}
\usepackage{bm}
\usepackage{color}
\usepackage{enumerate}

 \newif\ifpdf
 \ifx\pdfoutput\undefined
 \pdffalse
 \else
 \pdfoutput=1
 \pdftrue
 \fi
 \ifpdf
 \usepackage{graphicx}
 \usepackage{epstopdf}
 \else
 \usepackage{graphicx}
 \fi

\newcommand{\pt}{p_{\rm T}}

\begin{document}

\title{A hybrid model approach for strange and multi-strange hadrons in 2.76 A TeV Pb+Pb collisions}
\author{Xiangrong Zhu}
\affiliation{Department of Physics and State Key Laboratory of Nuclear Physics and Technology, Peking
University, Beijing 100871, China }

\author{Fanli Meng}
\affiliation{Department of Physics and State Key Laboratory of Nuclear Physics and Technology, Peking
University, Beijing 100871, China }

\author{Huichao Song}
\email[Correspond to\ ]{Huichaosong@pku.edu.cn}
\affiliation{Department of Physics and State Key Laboratory of Nuclear Physics and Technology, Peking
University, Beijing 100871, China}
\affiliation{Collaborative Innovation Center of Quantum Matter, Beijing 100871, China}
\affiliation{Center for High Energy Physics, Peking University, Beijing
100871, China}

\author{Yu-Xin Liu}
\affiliation{Department of Physics and State Key Laboratory of Nuclear Physics and Technology, Peking
University, Beijing 100871, China}
\affiliation{Collaborative Innovation Center of Quantum Matter, Beijing 100871, China}
\affiliation{Center for High Energy Physics, Peking University, Beijing
100871, China}

\begin{abstract}
Using the {\tt VISHNU} hybrid model, we calculate the multiplicity, spectra, and elliptic flow of $\Lambda$, $\Xi$ and $\Omega$ in 2.76 A TeV Pb+Pb collisions. Comparisons between our calculations and the ALICE measurements show that the model generally describes the soft hadron data of these strange and multi-strange hadrons at several centrality bins. Mass ordering of elliptic flow among $\pi$, K, p, $\Lambda$, $\Xi$ and $\Omega$ has also been studied and discussed. With a nice description of the particle yields, we explore chemical and thermal freeze-out of various hadrons species at the LHC within the framework of the {\tt VISHNU} hybrid model.
\end{abstract}

\pacs{25.75.-q, 12.38.Mh, 25.75.Ld, 24.10.Nz}
\maketitle

\section{INTRODUCTION}

Many measurements, such as jet quenching, elliptic flow, and valence quark number scaling of elliptic flow have provided strong evidences for the creation of the quark-gluon plasma (QGP) in heavy ion collisions at the Relativistic Heavy-Ion Collider (RHIC) and the Large Hadron Collider (LHC)~\cite{Rev-Arsene:2004fa,Gyulassy:2004vg,Muller:2006ee,Muller:2012zq}. With the formation of QGP and the restoration of chiral symmetry, strange and anti-strange quarks become abundant in the bulk medium above $T_c$, which subsequently enhance the productions of strange and multi-strange hadrons in
relativistic heavy ion collisions~\cite{Rafelski:1982pu}. In the past decades, different aspects of strange and multi-strange hadrons have been studied in theory ~\cite{Rafelski:1982pu,vanHecke:1998yu,Hamieh:2000tk,Letessier:2000ay,Heinz:1998st,Broniowski:2001uk,Huovinen:2001cy,Nonaka:2006yn,He:2011zx,Blume:2011sb} and in experiment~\cite{Adams:2003fy,Abelev:2007xp,Adams:2006ke,Adams:2005zg,Abelev:2013xaa,ABELEV:2013zaa,Abelev:2014pua}. It is generally believed that multi-strange hadrons, such as $\Xi$ and $\Omega$, directly carry the information of the QGP phase because of their small hadronic cross sections and the associated early decouplings from the system near $T_c$~\cite{vanHecke:1998yu}. Compared with common hadrons, their anisotropy flow are mainly developed in the QGP stage and less contaminated by the hadronic evolution.

Since the running of 2.76 A TeV Pb+Pb collisions at the LHC, the flow and other soft hadron data of all charged and identified hadrons have been studied by many groups within the framework of hydrodynamics~\cite{Schenke:2011tv,Qiu:2011hf,Song:2011qa,Song:2013qma,Song:2013tpa,Song:2012ua,Petersen:2011sb,Pang:2012he,Bozek:2012qs,Noronha-Hostler:2013gga}. Using the {\tt VISHNU} hybrid model~\cite{Song:2010aq} that connects (2+1)-d viscous hydrodynamics with a hadronic afterburner, we extracted the specific QGP shear viscosity $(\eta/s)_{QGP}$ from the elliptic data of all charged hadrons with MC-KLN initial conditions~\cite{Song:2011qa}. With the extracted value of $(\eta/s)_{QGP}$, {\tt VISHNU} provides a good description of the soft hadron data for $\pi$, K and p at the LHC~\cite{Song:2013qma}. Recently, the multiplicity, $\pt$-spectra and elliptic flow for $\Lambda$, $\Xi$ and $\Omega$ have been measured by the ALICE Collaboration~\cite{Abelev:2013xaa,ABELEV:2013zaa,Abelev:2014pua}. It is thus the right time to systematically study these strange and multi-strange hadrons at the LHC via the {\tt VISHNU} hybrid model.

This paper is organized as follows. In Sec.~II, we briefly introduce the {\tt VISHNU} hybrid model and its setup for
the calculations. Sec.~III compares our {\tt VISHNU} results with the ALICE measurements in 2.76 A TeV Pb+Pb collisions,
including the centrality dependence of the multiplicity density, $\pt$-spectra and differential elliptic flow for $\Lambda$,
$\Xi$ and $\Omega$. Sec.~IV studies and discusses mass ordering of elliptic flow among $\pi$, K, p, $\Lambda$, $\Xi$ and $\Omega$ at the LHC.
Sec.~V explores chemical and thermal freeze-out of various hadron species  during the {\tt UrQMD} evolution of {\tt VISHNU}.
Sec.~VI summarizes our current work and presents a brief outlook for the future.

\section{SETUP OF THE Calculation\label{sec:setup}}
In this section, we describe the inputs and setup for the {\tt VISHNU} calculations for the soft hadron data in 2.76 A TeV Pb+Pb collisions. The {\tt VISHNU} hybrid model~\cite{Song:2010aq} combines (2+1)-d relativistic viscous hydrodynamics ({\tt VISH2+1})~\cite{Song:2007fn} for the QGP fluid expansion with a microscopic hadronic transport model ({\tt UrQMD})~\cite{Bass:1998ca} for the hadron resonance gas evolution. The transition from hydrodynamics to the hadron cascade occurs on a switching hyper-surface with a constant temperature. Generally, the switching temperature $T_{sw}$ is set to 165 MeV which is close to the QCD phase transition temperature~\cite{Aoki:2006br,Borsanyi:2010bp,Bazavov:2011nk}. For the hydrodynamic evolution above $T_{\rm sw}$, we input an equation of state (EoS) constructed from recent lattice QCD data~\cite{Huovinen:2009yb,Shen:2010uy}.

Following Ref.~\cite{Song:2011qa,Song:2013qma,Song:2013tpa}, we input MC-KLN initial conditions~\cite{Drescher:2006ca,Hirano:2009ah} and start the hydrodynamic simulations at $\tau_0=0.9~{\rm fm}/c$. For computational efficiency, we implement single-shot simulations~\cite{Song:2010aq,Song:2011qa,Song:2013qma,Song:2013tpa,Song:2010mg} using smooth initial entropy density profiles generated by the MC-KLN model through averaging over a large number of events within specific centrality bins~\footnote{For recent development on event-by-event {\tt VISHNU} simulations, please refer to~\cite{Shen:2014vra,ChunPhDThesis}.}. Considering the conversion from total initial entropy to the final multiplicity of all charged hadrons, we cut the centrality bins through the distribution of total initial entropies obtained from the event-by-event fluctuating profiles from MC-KLN. Such centrality classification was once used by Shen in Ref.~\cite{ChunPhDThesis}, which is more close to the experimental cut from the measured multiplicity distribution. The normalization factor for the initial entropy density is fixed by the charged hadron multiplicity density in the most central collisions ($dN_{\rm ch}/d\eta \approx 1601\pm60$ from ALICE~\cite{Aamodt:2010cz}). The $\lambda$ parameter in the MC-KLN model, which quantifies the gluon saturation scale in the initial gluon distributions~\cite{Drescher:2006ca}, is tuned to $0.138$ for a better fit of the centrality dependent multiplicity density for all charged hadrons.

The QGP specific shear viscosity $(\eta/s)_{QGP}$ is set to 0.16 for MC-KLN initial conditions. Such combined setting for {\tt VISHNU} once nicely described the elliptic flow of pions, kaons and protons in 2.76 A TeV Pb+Pb collisions~\cite{Song:2013qma}. Here, we continue to use it to further study the soft hadron data of strange and multi-strange hadrons at the LHC. To simplify the theoretical investigations, we set the bulk viscosity to zero and neglect the net baryon density and heat conductivity for the QGP systems created at the LHC.

\section{MULTIPLICITY, SPECTRA, AND ELLIPTIC FLOW for $\Lambda$, $\Xi$ and $\Omega$ \label{sec:comResults}}

The multiplicity, spectra and elliptic flow of pions, kaons and protons in 2.76 A TeV Pb+Pb collisions have been studied in our early paper~\cite{Song:2013qma}. We showed that, with MC-KLN initial conditions, $\eta/s=0.16$ and other parameters fixed from the related data of all charged hadrons, {\tt VISHNU} could nicely describe the soft hadron data of pions, kaons and protons at the LHC. We also found that baryon-antibaryon (${\it B}$-$\bar{B}$) annihilations in the {\tt UrQMD} module of {\tt VISHNU} could reduce the proton yields by O(30\%), leading to nice fits of the proton data measured by ALICE. In this section, we extend our early {\tt VISHNU} simulations to high-statistics runs to further study the soft hadron data for the strange and multi-strange hadrons $\Lambda$, $\Xi$ and $\Omega$ in 2.76 A TeV Pb+Pb collisions~\footnote{The $\phi$ meson is another important multi-strange hadron that might directly carry the QGP information due to its small hadronic cross-sections. {\tt VISHNU} predictions for the spectra and elliptic flow of $\phi$ can be found in Ref.~\cite{Song:2013qma}. However, later comparisons showed pretty large deviations between theory and experiment~\cite{Abelev:2014pua}. Unlike other hadrons, $\phi$ mesons are mainly reconstructed from the strong decay channel $\phi\rightarrow K^+ K^-$, rather than being directly measured. The succeeding hadronic scatterings of kaons might contaminate the weak signals of $\phi$. In this paper, we will not show and discuss the results of $\phi$, but just quickly mention our early work~\cite{Song:2013qma} and leave the puzzle of $\phi$ for future study.}.

\begin{figure}[t]
 \includegraphics[width=0.9\linewidth,height=10cm]{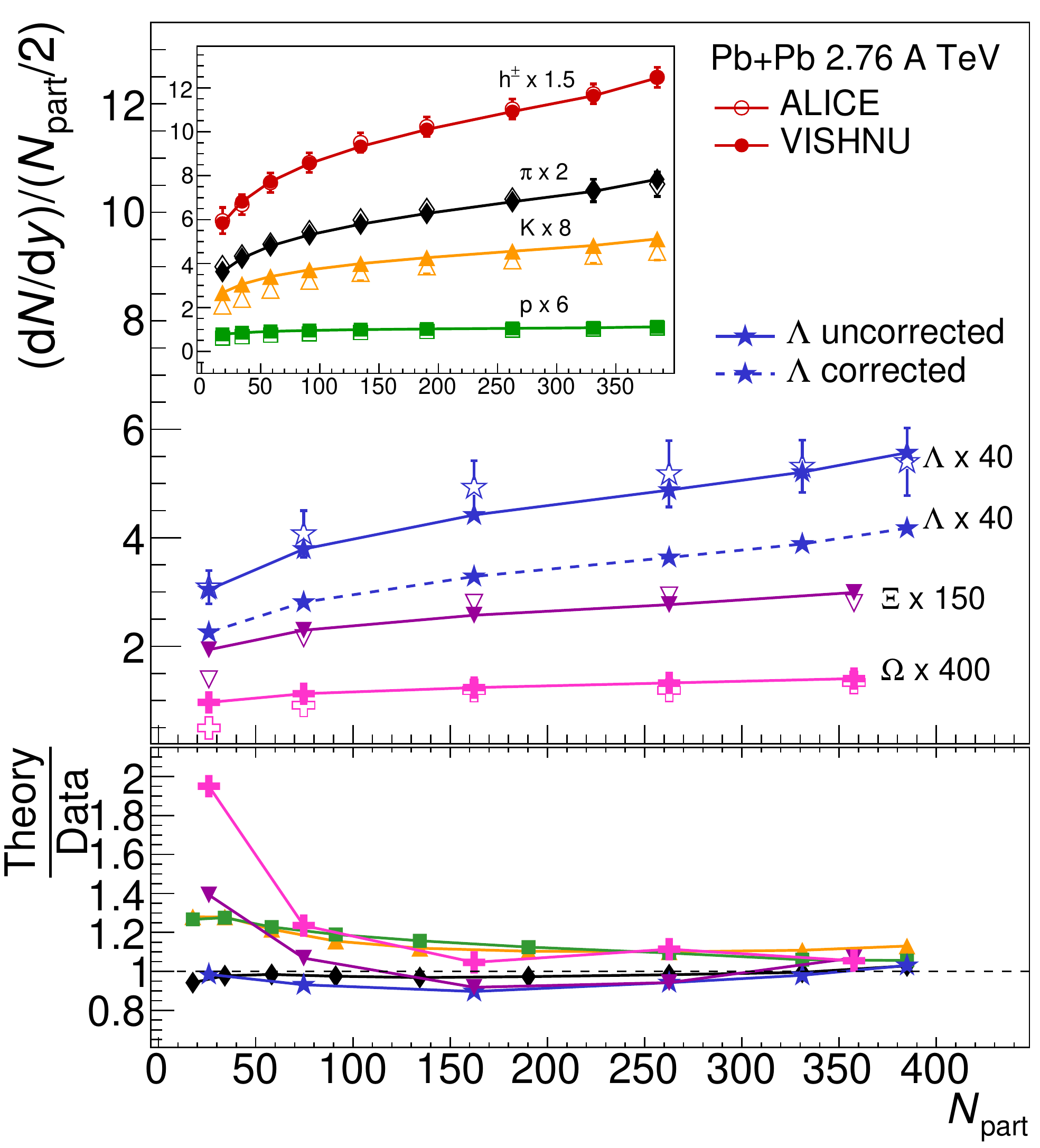}
 \caption{(Color online) Centrality dependence of the multiplicity density per participant pair,
 $(dN/dy)/(N_{\rm part}/2)$ for $\Lambda$, $\Xi$  and
 $\Omega$ in 2.76 A TeV Pb+Pb collisions. Insert: $(dN/dy)/(N_{\rm part}/2)$ for $\pi$, K, p and for all charged hadrons. Experimental data are from the ALICE Collaboration~\cite{Aamodt:2010cz, Abelev:2013xaa, Abelev:2013vea, ABELEV:2013zaa}.
 Theoretical curves are calculated with the {\tt VISHNU} hybrid model, using MC-KLN initial conditions, $\eta/s=0.16$ and $T_{sw}=165 \ \mathrm{MeV}$.}.
\end{figure}
\begin{figure*}[t]
  \includegraphics[width=0.8\linewidth,height=9cm,clip=]{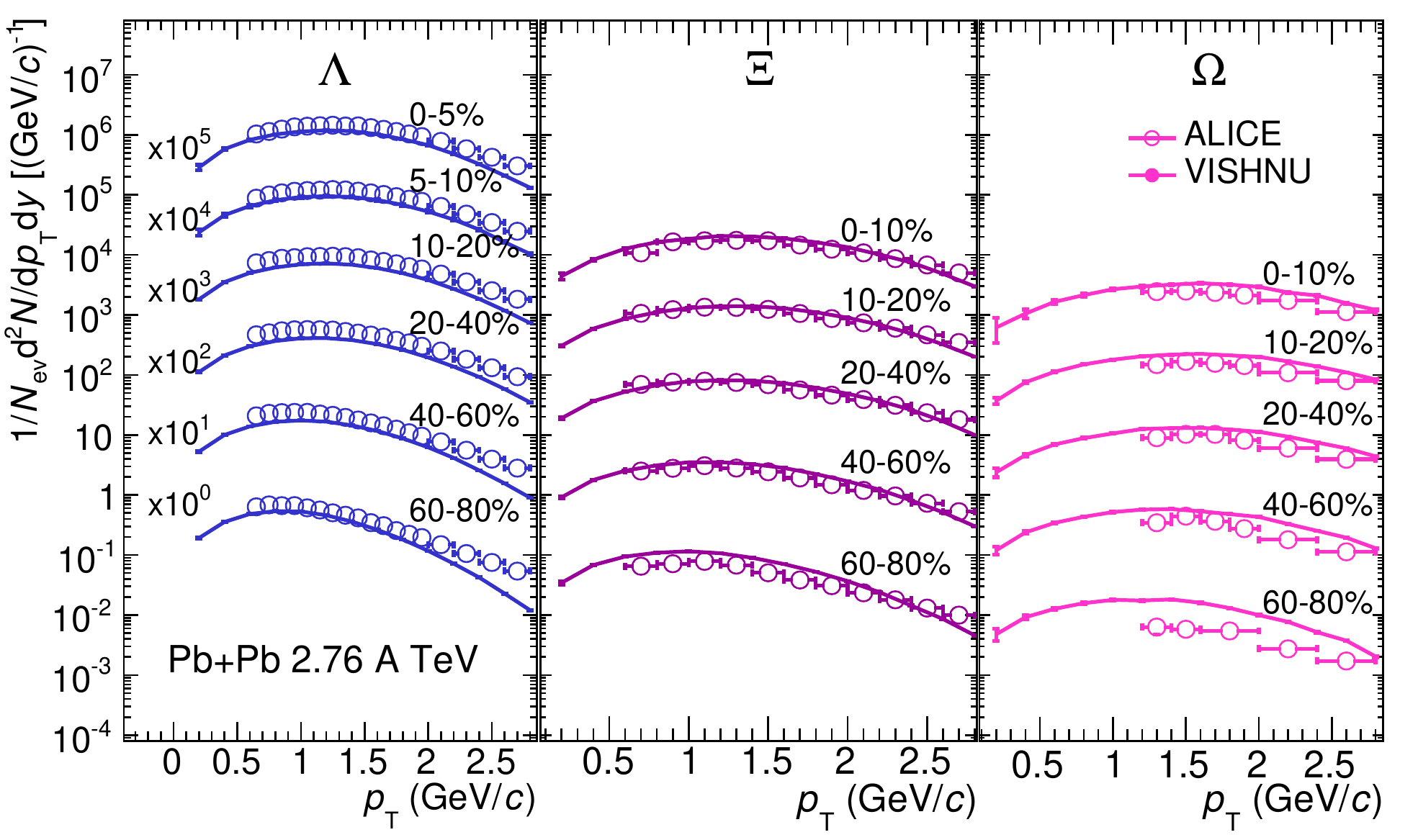}
  \caption{(Color online) Transverse momentum spectra of $\Lambda$, $\Xi$ and $\Omega$ at various centralities in 2.76 A TeV Pb+Pb collisions. Experimental data  are taken from ALICE~\cite{Abelev:2013xaa, ABELEV:2013zaa}. Theoretical curves are calculated with {\tt VISHNU} with the same inputs as for Fig.1. From top to bottom the curves correspond to  0-10\% ($\times10^{4}$), 10-20\% ($\times10^{3}$), 20-40\% ($\times10^{2}$), 40-60\% ($\times10^{1}$) and 60-80\% ($\times$1) centrality, respectively, where the factors in parentheses are the multipliers applied to the spectra for clear separation. Spectra of $\Lambda$ start from 0-5\% ($\times10^{5}$) and 5-10\% ($\times10^{4}$), instead of 0-10\%.
 \label{fig:PtLXO}}
\end{figure*}

Figure 1 shows the centrality dependence of the multiplicity density per participant pair $(dN/dy)/(N_{\rm part}/2)$ for $\Lambda$, $\Xi$ and $\Omega$ in 2.76 A TeV Pb+Pb collisions~\footnote{We notice that the measured multiplicity of $\Lambda$ from ALICE are contaminated by the feed-down decays of $\Sigma^{0}$ and $\Sigma$(1385)~\cite{Abelev:2013xaa}. However, the {\tt UrQMD} module of {\tt VISHNU} only includes strong resonance decays, but without any weak decays. To partially account the effects from weak decays, e.g. $\Sigma^{0}\rightarrow\Lambda+\gamma$, we directly sum the multiplicity of $\Lambda$ and $\Sigma^{0}$ from {\tt VISHNU} to get a corrected curve of $\Lambda$ (the solid blue line with star symbols). The original yields of $\Lambda$ from {\tt VISHNU} are also shown in Fig.~1, which is presented by the dashed blue line with star symbols. In our estimations, the $\Sigma^{0}\rightarrow\Lambda+\gamma$ channel contributes $\sim$ 30\% additional $\Lambda$ productions.}. In the insert, we plot the corresponding curves for pions, kaons, protons and for all charged hadrons that were once presented in our early paper~\cite{Song:2013qma} for the easiness of reference.  The inputs of our current calculations are close to the ones used in~\cite{Song:2013qma,Song:2013tpa}, except for two points: 1) changing the $\lambda$ parameter in the MC-KLN model from 0.28 to 0.138, 2) cutting the centrality bins through initial entropy rather than the participant number $N_{part}$  (please refer to Sec.~II for details). Compared with the early setup, these two changes mainly improve the description of the centrality dependent multiplicity for all charged hadrons and for pions, but they have small influence on other theoretical results, such as the elliptic flow of all charged and identified hadrons, etc..

One finds that, {\tt VISHNU} nicely describes these $(dN/dy)/(N_{\rm part}/2)$ curves for all investigated hadrons. Like the case of protons, ${\it B}$-$\bar{B}$ annihilations also reduce the yields of strange and multi-strange baryons with O(30\%) for $\Lambda$, O(20\%) for $\Xi$ and $\Omega$ in the most central Pb+Pb collisions(please refer to Fig.~5 in sec.~V). The lower panel of Fig.~1 shows the difference between the theoretical calculated and the experimental measured particle yields. From the most central to semi-peripheral collisions, the deviations are all within 20\%. For the 60-80\% centrality bin, the differences increase to 40\% for $\Xi$, and 100\% for $\Omega$. This indicates that the strangeness no longer reach chemical equilibrium in the small system created in peripheral Pb+Pb collisions.

\begin{figure*}[tbph]
 \includegraphics[width=0.9\linewidth, height=6cm]{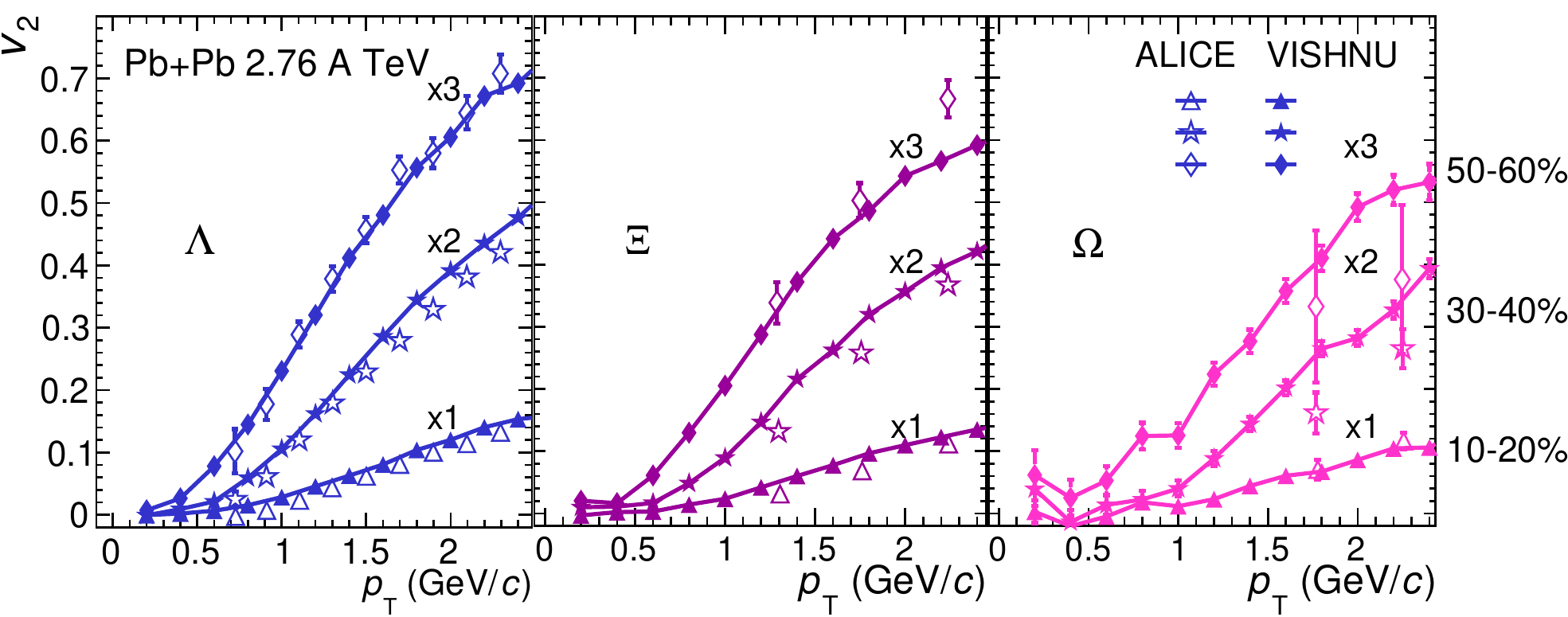}
  \caption{(Color online) Differential elliptic flow of strange hadrons $\Lambda$, multi-strange hadrons $\Xi$
  and $\Omega$ at 10-20\%,  30-40\% and 50-60\% centralities in 2.76 A TeV Pb+Pb collisions.
  Experimental data are from ALICE~\cite{Abelev:2014pua}, theoretical curves are calculated from {\tt VISHNU} with the same inputs as for Fig.~1 and Fig.~2.
 \label{fig:V2LXO}}
\end{figure*}
\begin{figure*}[tph]
 \includegraphics[width=0.6\linewidth]{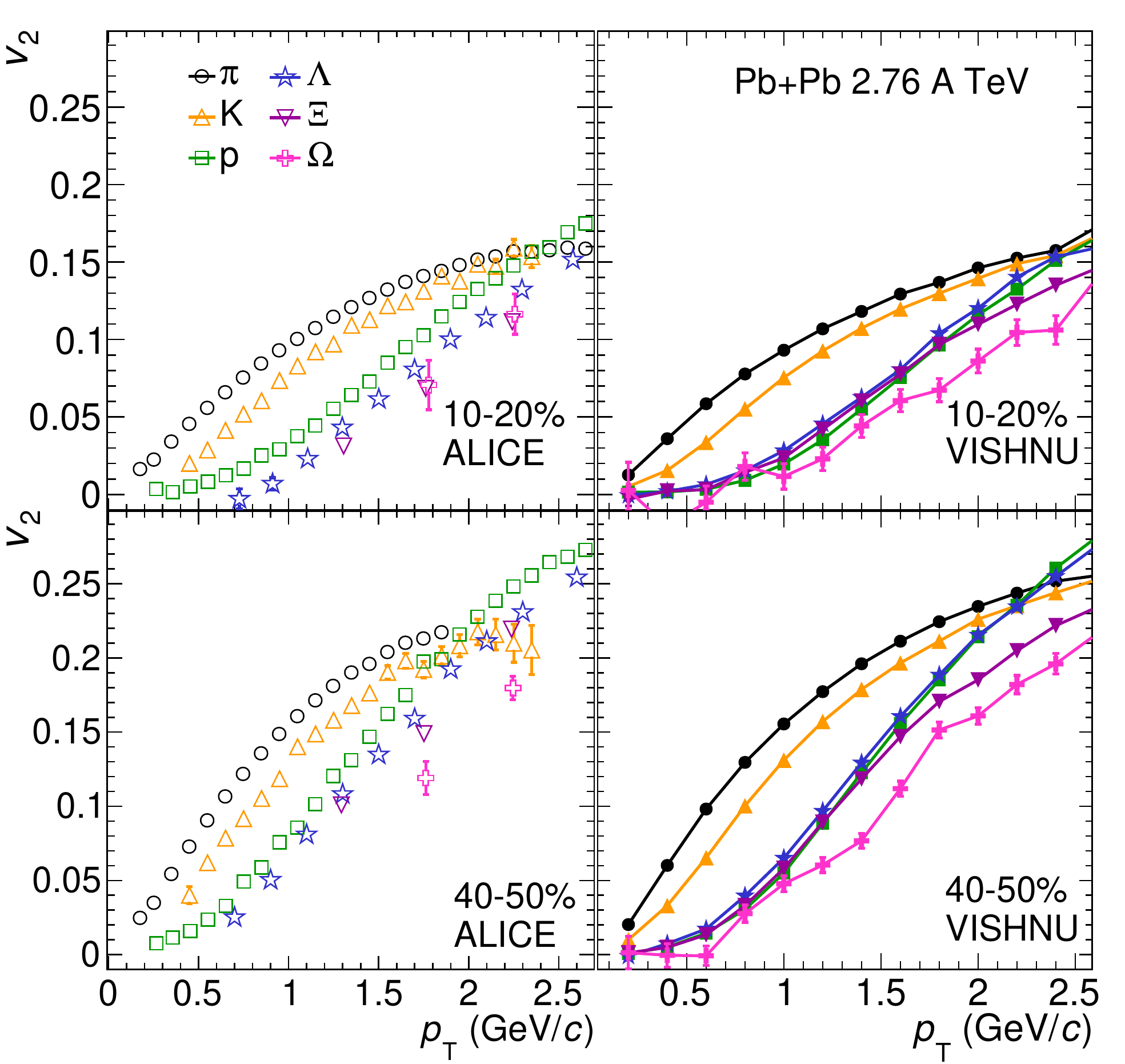}
  \caption{(Color online) Differential elliptic flow of $\pi$, K, p, $\Lambda$,
  $\Xi$ and $\Omega$ at 10-20\% and 40-50\% centralities in 2.76 A TeV Pb+Pb collisions. Left panels (a) and (c) are plotted with $v_2$ from ALICE~\cite{Abelev:2014pua}, right panels (b) and (d) are plotted with $v_2$ from {\tt VISHNU}.
 \label{fig:Huovinen:2001cy}}
\end{figure*}

In Fig.~2, we compare the transverse momentum spectra of $\Lambda$, $\Xi$ and $\Omega$ from {\tt VISHNU} with the measurements from ALICE. In general, {\tt VISHNU} describes the $p_T$ spectra of these strange and multi-strange hadrons from the most central to semi-peripheral collisions, except for the 60-80\% centrality bin. Here, the theoretical curves of $\Lambda$ are plotted with the original values from {\tt VISHNU} without weak decays. As a result, they
are about 30\% lower than the experimental measurements with weak decay contaminations. The $\Omega$ spectra from {\tt VISHNU} are slightly higher than the experimental data for most centralities, but obviously above the data at the 60-80\% centrality bin. Such deviations between theory and experiment are consistent with the model and data differences for the centrality dependent multiplicity shown in Fig.~1. In spite of the normalization issues, {\tt VISHNU} nicely fits the slope of the $p_T$ spectra for $\Lambda$, $\Xi$ and $\Omega$ at various centralities. Together with the early nice descriptions of the $p_T$ spectra for pions, kaons and protons~\cite{Song:2013qma}, it indicates that {\tt VISHNU} generates a proper amount of radial flow, during its QGP and hadronic evolution, to push the spectra of various hadrons.

Figure~3 presents the differential elliptic flow of $\Lambda$, $\Xi$ and $\Omega$ at three chosen centralities in 2.76 A TeV Pb+Pb collisions. The experimental data are from ALICE, which are measured with the scalar product method~\cite{Abelev:2014pua}. The theoretical lines are calculated from {\tt VISHNU} with MC-KLN initial conditions and $(\eta/s)_{\tt QGP}=0.16$. Such inputs once nicely described the elliptic flow data of pions, kaons and protons at the LHC~\cite{Song:2013qma}. In principle, the current calculations can be considered as extensions of the early simulations~\cite{Song:2013qma}. Fig.~3 shows that the elliptic flow data below 2 GeV for $\Lambda$, $\Xi$ and $\Omega$ are fairly described by {\tt VISHNU} within the statistical error bars. Above 2 GeV, the descriptions of the elliptic flow for $\Xi$ at 50-60\% and for $\Omega$ at 30-40\% and 50-60\% become worse. On the other hand, viscous corrections probably become too large in that higher $p_T$ region, making the hydrodynamic description in {\tt VISHNU} lost its predictive power.

\section{Mass ordering of elliptic flow}

Mass ordering of elliptic flow among various hadron species reflects the interaction between the radial and elliptic flow during the hadronic evolution. The radial flow tends to push the heavier particles at lower $\pt$ to higher $\pt$, leading to a mass ordering of the $p_T$ dependent elliptic flow below 1.5-2 GeV that decreases with the increase of hadron mass. Such $v_2$ mass-ordering has been discovered in experiments at both RHIC and the LHC~\cite{Adams:2004bi,Abelev:2014pua,Abelev:2010tr,Snellings:2014vqa}, which has also been studied within the framework of hydrodynamics~\cite{Huovinen:2001cy,Hirano:2007ei,Shen:2011eg,Bozek:2009dw,Song:2013tpa} and the blast wave model~\cite{Adler:2001nb,Retiere:2003kf}.

In Fig.~4, we investigate mass ordering of elliptic flow among $\pi$, K, p, $\Lambda$, $\Xi$ and $\Omega$  in 2.76 A TeV Pb+Pb collisions. For clear presentations, the ALICE data and the {\tt VISHNU} results are plotted in separate panels for the two chosen centralities at 10-20\% and
40-50\%. Calculations in~\cite{Song:2013qma} and in this paper (Fig.~3) have respectively showed that {\tt VISHNU} generally describes $v_2(p_T)$ for various individual hadrons over a wide range of centralities. However, further comparisons in Fig.~4 illustrate that {\tt VISHNU} could not describe the $v_2$ mass ordering among all hadron species. Although {\tt VISHNU} nicely describes the mass ordering among $\pi$, K, p, and $\Omega$, it fails to correctly describe the mass ordering among  p, $\Lambda$ and $\Xi$. In contrast, pure viscous hydrodynamics {\tt VISH2+1} has correctly predicted the relative mass-ordering among these investigated hadrons, but it has difficulties to roughly fit the $v_2$ data for these heavier hadrons like p, $\Lambda$, $\Xi$ at the 10-20\% centrality bin~\cite{Shen:2011eg}.

Compared with the elliptic flow of individual hadrons, the $v_2$  mass-splittings between different hadron species reveal more details for the hadronic
evolution. Although {\tt VISHNU} could improve the description of $v_2(p_T)$ for the hadron species like p, $\Lambda$ and $\Xi$ through its microscopic hadronic scatterings, it slightly under-predicts the proton $v_2$ below 2 GeV, leading to inverse $v_2$ mass ordering between p and $\Lambda$, and accidental overlaps of the elliptic flow for p and $\Xi$  below 1.5 GeV. An initial flow could enhance the radial flow in the hadronic stage, which is thus expected to improve the description of $v_2$ mass ordering within the framework of the hybrid model. Meanwhile, the {\tt UrQMD} hadronic cross sections also need to be re-evaluated and improved. These have not been done currently and should be investigated in the near future.

\section{Chemical and thermal freeze-out of various hadron species \label{sec:chemicalFreezeout}}
In this section, we investigate chemical and thermal freeze-out of various hadron species during the {\tt UrQMD} evolution within the framework of the {\tt VISHNU} hybrid model.

During the QGP fireball evolution, a large number of hadrons are produced near $T_c$, which subsequently undergo inelastic and elastic collisions in the hadronic phase. With the termination of inelastic collisions, the yields of each hadron species no longer change. The system is considered to reach chemical freeze-out. Thermal freeze-out happens later, which is associated with the end of elastic collisions. After that, the momentum distributions of final produced hadrons are fixed.

In the statistical model, the chemical freeze-out temperature $T_{ch}$ and the baryon chemical potential $\mu_{b}$ are extracted from the particle yields of various hadrons~\cite{BraunMunzinger:2001ip,BraunMunzinger:2003zd,Letessier:2005qe,Becattini:2005xt}. A systematic study of the related data at top RHIC energy gives $T_{ch}\simeq165 \ \mathrm{MeV}$~\cite{BraunMunzinger:2001ip}. This temperature could describe the yields of many identified hadrons in 2.76 A TeV Pb+Pb collisions, but obviously over-predicts the protons/antiprotons data at the LHC. A good description of the $p/\bar{p}$ data requires a lower chemical freeze-out temperature around $150 \ \mathrm{MeV}$. However, such lower temperature breaks the early nice description of $\Xi$ and $\Omega$ yields once achieved with $165 \ \mathrm{MeV}$~\cite{Becattini:2012xb,Stachel:2013zma}.

To study the above proton puzzle from the statistical model, we systematically investigated the soft hadron data for $\pi$, K, p at both RHIC and the LHC with the {\tt VISHNU} hybrid model~\cite{Song:2013qma}. We found that baryon and anti-baryon annihilations influence the transport of protons/antiprotons during the hadronic evolution, leading to a largely improved description of the $p/\bar{p}$ yields when compared with the case without ${\it B}$-$\bar{B}$ annihilations~\footnote{Other related work could be found in Ref.~\cite{Becattini:2012xb,Becattini:2012sq,Becattini:2014hla}.}. Meanwhile, other soft hadron data of $\pi$, K and p are also nicely fitted in general. This paper extends the early {\tt VISHNU} calculations to further study strange and multi-strange hadrons at the LHC. Sec.~III has showed a nice description of the paricle yields for $\Lambda$, $\Xi$ and $\Omega$, together with good fits of the spectra and elliptic flow for these hadrons.

In our calculations, the switching temperature, that connects the hydrodynamic description of the QGP expansion to the Boltzmann approach for the hadron resonance gas evolution, is set to 165 MeV at both RHIC and the LHC. However, this temperature can not be identified as the chemical freeze-out temperature in the statistical model, since ${\it B}$-$\bar{B}$ annihilations and other inelastic collisions are still frequent during the early hadronic evolution, which constantly change the yields of various hadrons. Instead, different hadronic scatterings in {\tt UrQMD} lead to a hadron species dependent
chemical freeze-out procedure.

Figure~5 studies the time evolution of particle yield density for $\pi$, K, p, $\Lambda$, $\Xi$ and $\Omega$ during the {\tt UrQMD} hadronic expansion. This investigation is still done within the {\tt VISHNU} simulations for 2.76 A TeV Pb+Pb collisions, but exports the {\tt UrQMD} intermediate results at different evolution times. For the easiness of comparison, Left panels (a)-(f) plot the time evolution of relative particle yield density: $\frac{dN}{dy}(t)/\frac{dN}{dy}(0)$. Here, $\frac{dN}{dy}(t)$ and $\frac{dN}{dy}(0)$ denote the particle yield density at mid-rapidity at later evolution time and at the starting time, respectively.

For the simulations without ${\it B}$-$\bar{B}$ annihilations, the yields of $\Xi$ and $\Omega$ almost do not change. This indicates that these two multi-strange baryons experience early chemical freeze-out near the switching hyper-surface of {\tt VISHNU}. For other hadron species, their yields constantly change with the {\tt UrQMD} evolution. By the end of the evolution, the yields of K and p respectively decrease $\sim 5\%$ and $\sim 10\%$, and the yield of $\Lambda$ increases $\sim 40\%$. Meanwhile, the changing rates for the particle yield density of K, p and $\Lambda$ show wide peaks along the time axis (Panel (g)), illustrating that the associated inelastic collisions are still frequent after 10-20 ${\rm fm}/c$. This indicates these hadrons experience later chemical freeze-out. The yield of pions only slightly increases during the {\tt UrQMD} evolution without ${\it B}$-$\bar{B}$ annihilations. However, this is not necessarily associated with early chemical freeze-out of pions. Instead, pions maintain relative chemical equilibrium below $T_c$ through frequent quasi-elastic collisions, e.g., $\pi\pi\leftrightarrow\rho$, $\pi N \leftrightarrow\Delta$, etc.~\cite{Hirano:2002ds,Teaney:2002aj,Kolb:2002ve,Huovinen:2007xh}.

\begin{figure}[t]
 \includegraphics[width=0.99\linewidth,height=10.5cm]{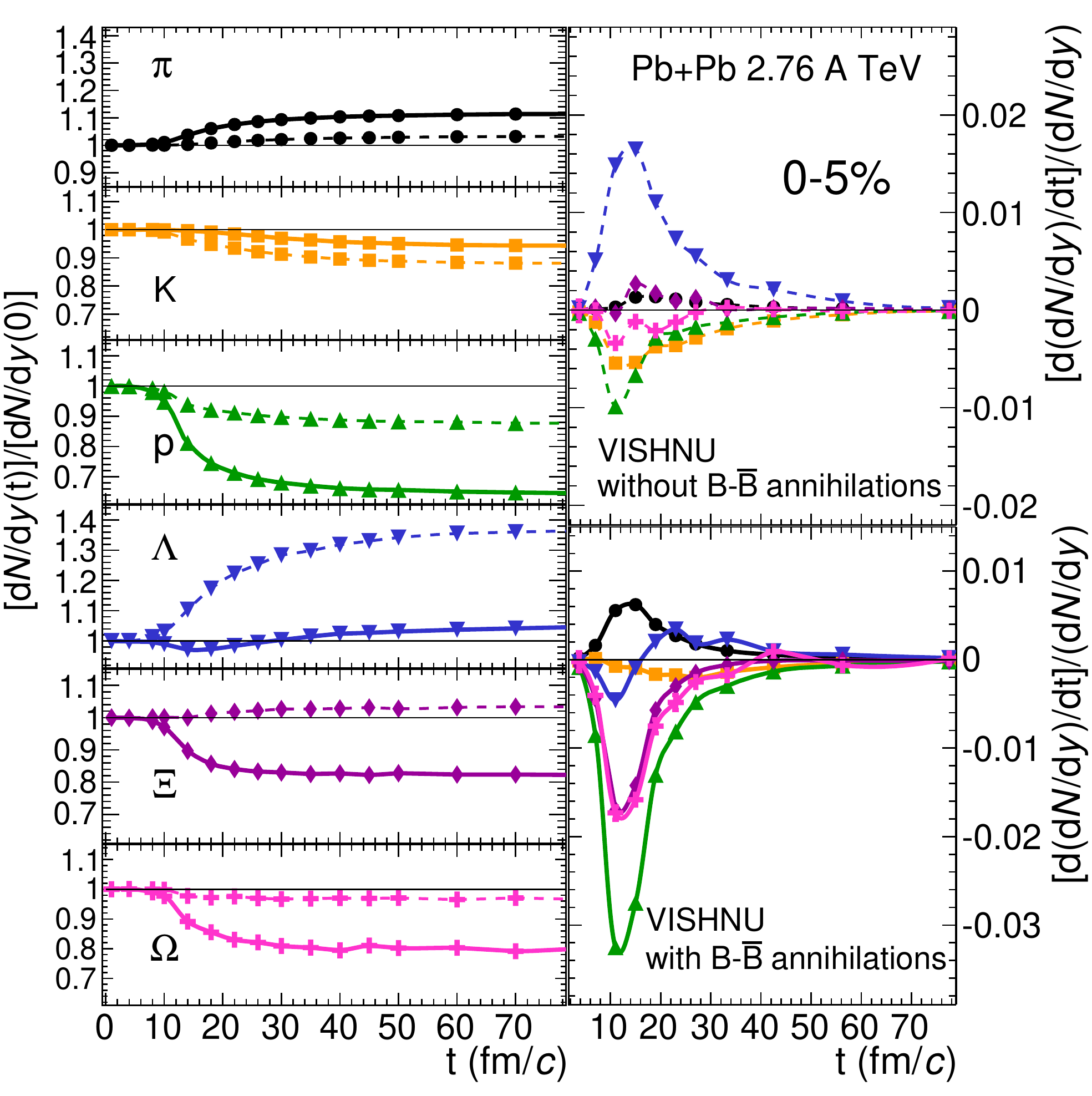}
  \caption{(Color online)  Left panels (a)-(f): time evolution of relative particle yield density $\frac{dN}{dy}(t)/\frac{dN}{dy}(0)$ for $\pi$, K, p, $\Lambda$, $\Xi$ and $\Omega$ during the {\tt UrQMD} expansion of {\tt VISHNU}. Right panels (g) and (h): time evolution of the changing rate for the corresponding particle yield density. Solid/dashed lines denote the {\tt VISHNU} simulations with/without ${\it B}$-$\bar{B}$ annihilations in the most central 2.76 A TeV Pb+Pb collisions.
 \label{fig:timedNdy}}
\end{figure}
\begin{figure}[t]
 \includegraphics[width=0.99\linewidth,height=10.5cm]{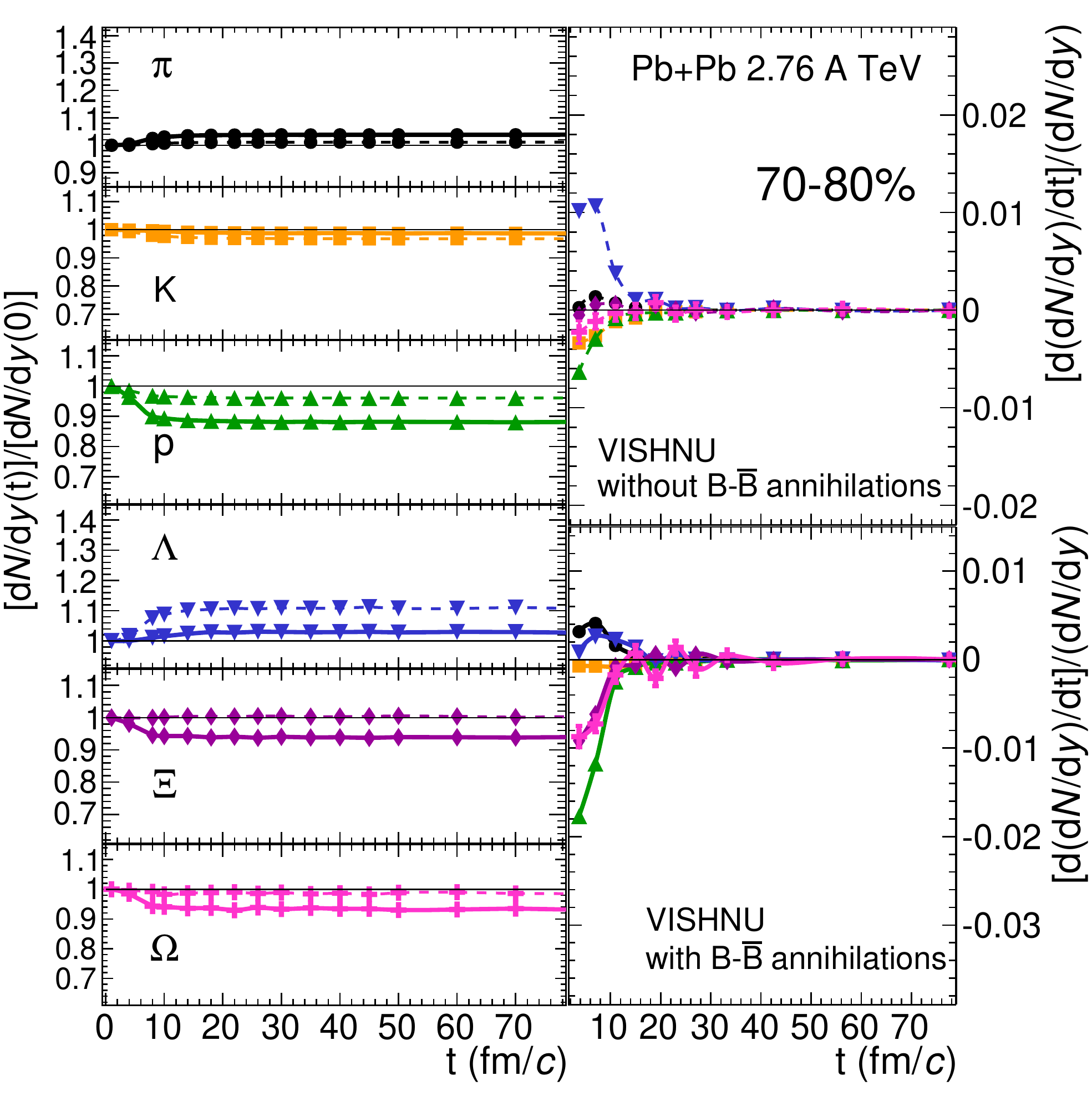}
  \caption{(Color online) Similar to Fig.~5, but for 70-80\% centrality bin.
 \label{fig:timedNdy}}
\end{figure}
\begin{figure*}[t]
 \includegraphics[width=0.8\linewidth,height=8.7cm]{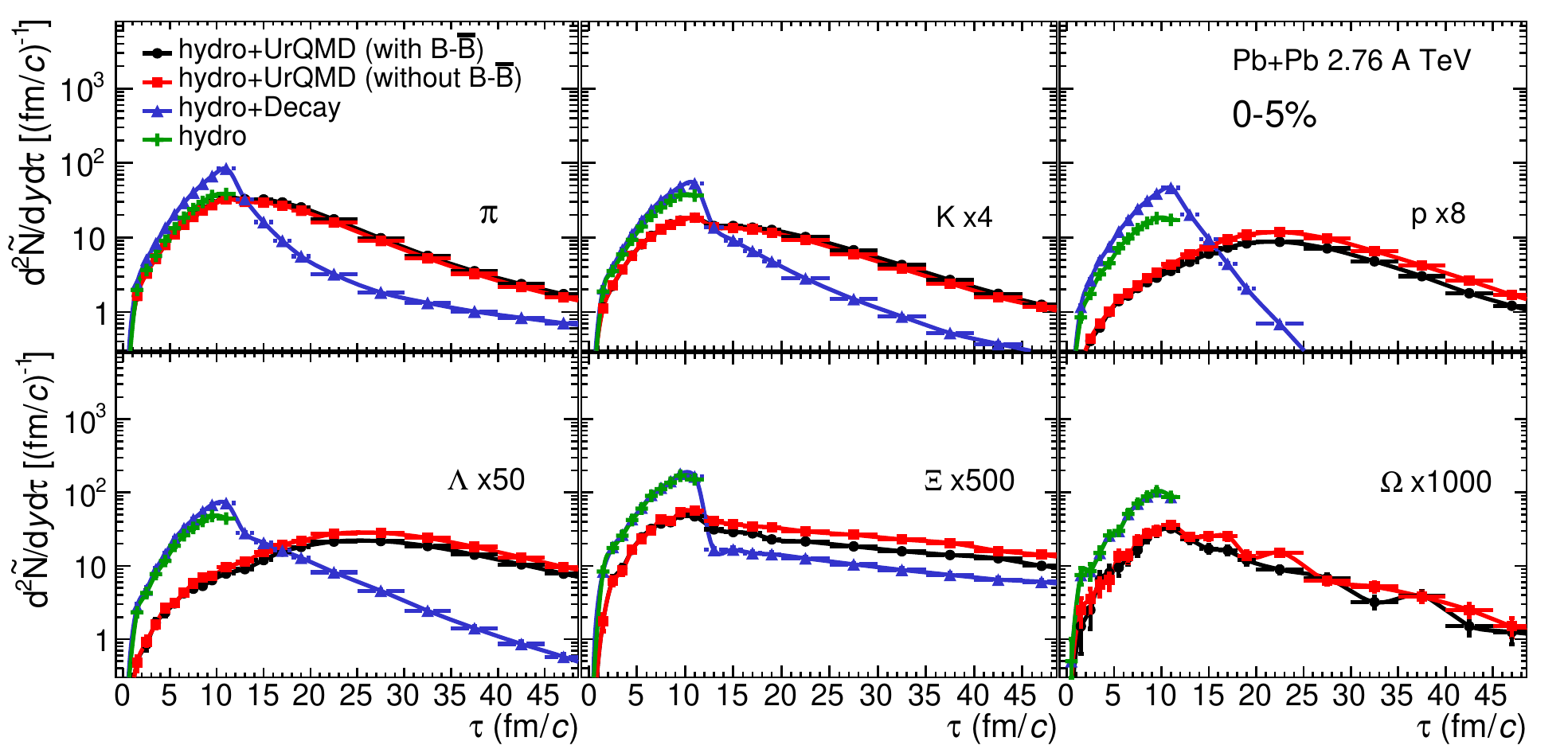}
  \caption{(Color online) Thermal freeze-out time distributions for $\pi$, K, p, $\Lambda$, $\Xi$ and $\Omega$
  in the most central Pb+Pb collisions, calculated from hydrodynamics (green), hydrodynamics+resonance decay (blue),
  {\tt VISNU} without ${\it B}$-$\bar{B}$ annihilations (red) and with ${\it B}$-$\bar{B}$ annihilations (black).
 \label{fig:fourmodel}}
\end{figure*}

The ${\it B}$-$\bar{B}$ annihilations ($p\bar{p}\rightarrow n\pi$, etc.) mainly influence the baryon's transport in {\tt UrQMD}, leading to $\sim 30\%$ reductions for the p and $\Lambda$ yields, $\sim 20\%$ reductions for the $\Xi$ and $\Omega$ yields in the most central Pb+Pb collisions. Meanwhile, the yields of $\pi$ and K slightly increase by $\sim 5\%$ through the annihilation channels. Although these two multi-strange hadrons, $\Xi$ and $\Omega$, rarely interact with other hadrons during the {\tt UrQMD} evolution, the annihilations with their own anti-particles delay their chemical freeze-out. This is presented by the wide peaks on the changing rate curves for these two multi-strange baryons in panel (h). But, compared with other curves such as the proton one, $\Xi$ and $\Omega$ still experience early chemical freeze-out. The ${\it B}$-$\bar{B}$ annihilations almost balance with the other inelastic collision channels on the production of $\Lambda$ and K. As a result, the yields of these two hadrons only slightly change during the hadronic evolution.

Fig.~6 is similar to Fig.~5, but for the 70-80\% centrality bin. For the case without ${\it B}$-$\bar{B}$ annihilations, the particle yields of various hadrons almost do not change during the {\tt UrQMD} evolution. Compared with the most central Pb+Pb collisions, the number of in-elastic collisions in the hadronic phase
are greatly reduced. Fig.~6 also showed that the ${\it B}$-$\bar{B}$ annihilations decrease the baryon yields for p, $\Lambda$, $\Xi$ and $\Omega$ by 5-8\% in peripheral collisions, but most of the annihilations happen before 10 ${\rm fm}/c$.

Shortly speaking, Fig.~5 and Fig.~6 mainly concentrate on studying time evolution of various hadron yields, which indirectly reflect the inelastic collisions in {\tt UrQMD}. A further analysis of the space-time distributions of the last inelastic collisions will reveal direct information for chemical freeze-out, which may even help us to extract effective chemical freeze-out temperatures of various hadron species. Unfortunately, the current version of {\tt UrQMD} does not record such intermediate information. We have to leave it for future study.

Besides 4-momentum of final produced hadrons, {\tt UrQMD} also outputs the positions (in space and time) of the last elastic collisions or resonance decays that directly reflect thermal freeze-out of the evolving system. Here we define the time distributions of the last collisions or decays for various hadrons species as the corresponding \emph{thermal freeze-out time distributions}. On the other hand, they can also be considered as the production-time distributions of specific hadron species during the {\tt UrQMD} evolution.

Figure~7 shows thermal freeze-out time distributions for $\pi$, K, p, $\Lambda$, $\Xi$ and $\Omega$ in the most central Pb+Pb collisions. To study the hadronic scattering effects, we set four comparison simulations: \textbf{1)} viscous hydrodynamics terminated at $T_{sw}=165\ \mathrm{MeV} $ with only thermal hadron emissions, \textbf{2)} viscous hydrodynamics terminated at $T_{sw}$ with thermal hadron productions and succeeding resonance decays, \textbf{3)} {\tt VISHNU} without ${\it B}$-$\bar{B}$ annihilations, \textbf{4)} {\tt VISHNU} with ${\it B}$-$\bar{B}$ annihilations. Here, both viscous hydrodynamics and {\tt VISHNU}
simulations input the same initial conditions, EoS, and other related parameters as described in Sec.~II.

The thermal freeze-out time distributions for various hadrons in case \textbf{(1)} all stop around 10 ${\rm fm}/c$, because the hydrodynamic evolution terminates around that time in the most central collisions. Comparing the thermal freeze-out time distributions from hydrodynamics (case \textbf{1}) and from  hydrodynamics+resonance decays (case \textbf{2}), we find a certain portion of the resonance decays happen near the hydrodynamic freeze-out surface, which largely enhance the productions of pions and protons before 10 ${\rm fm}/c$. Meanwhile, the long-lived resonances also contribute later hadron productions after 10 ${\rm fm}/c$, which results in long tails for the distribution curves of $\pi$, K, p, $\Lambda$ and $\Xi$.  We notice that there is no change for the $\Omega$ curves between case \textbf{(1)} and case \textbf{(2)}. {\tt UrQMD} only includes hadrons below 2 GeV, the associated resonance decays do not contribute to the production of this heavy multi-strange baryon.

The {\tt UrQMD} hadronic scatterings in case \textbf{(3)} and \textbf{(4)} broaden thermal freeze-out time distributions of all hadron species, which shift the averaged hadron production times before 10 ${\rm fm}/c$ in case \textbf{(1)} and \textbf{(2)} to later values ranging from 10 ${\rm fm}/c$ to 40 ${\rm fm}/c$. We also observe that the ${\it B}$-$\bar{B}$ annihilations further decrease the productions of p, $\Lambda$, $\Xi$ and $\Omega$ as shown in Fig.~5. In general, such annihilations do not change the shape of these thermal freeze-out time curves.

Figure~8 compares the thermal freeze-out time distributions from the most central collisions with the ones from peripheral collisions. Here, the results are from the {\tt VISHNU} simulations with ${\it B}$-$\bar{B}$ annihilations. We find that the peaks of the thermal freeze-out time distributions for $\pi$, K, p, $\Lambda$, $\Xi$ and $\Omega$ are all shifted to much earlier time in peripheral collisions, because the created QGP fireball there has much smaller volume and shorter lifetime.

\begin{figure*}[t]
 \includegraphics[width=0.8\linewidth,height=8.7cm]{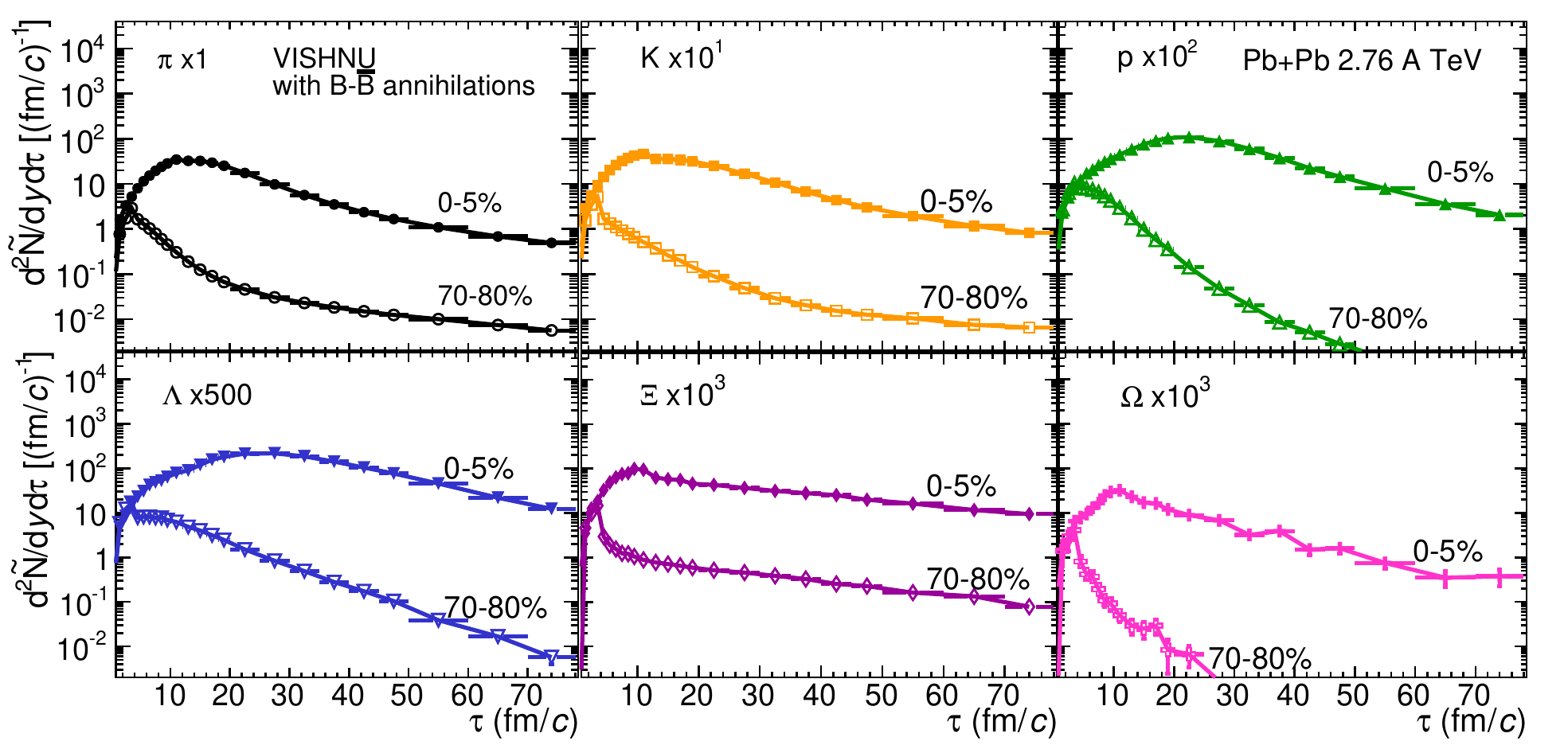}
  \caption{(Color online) Comparisons of the thermal freeze-out time distributions for $\pi$, K, p,
  $\Lambda$, $\Xi$ and $\Omega$  between the most central and peripheral Pb+Pb collisions, calculated from {\tt VISHNU}
  with ${\it B}$-$\bar{B}$ annihilations.
 \label{fig:WBB05a7080}}
\end{figure*}

\begin{figure}[tph]
 \includegraphics[width=0.85\linewidth,height=9cm]{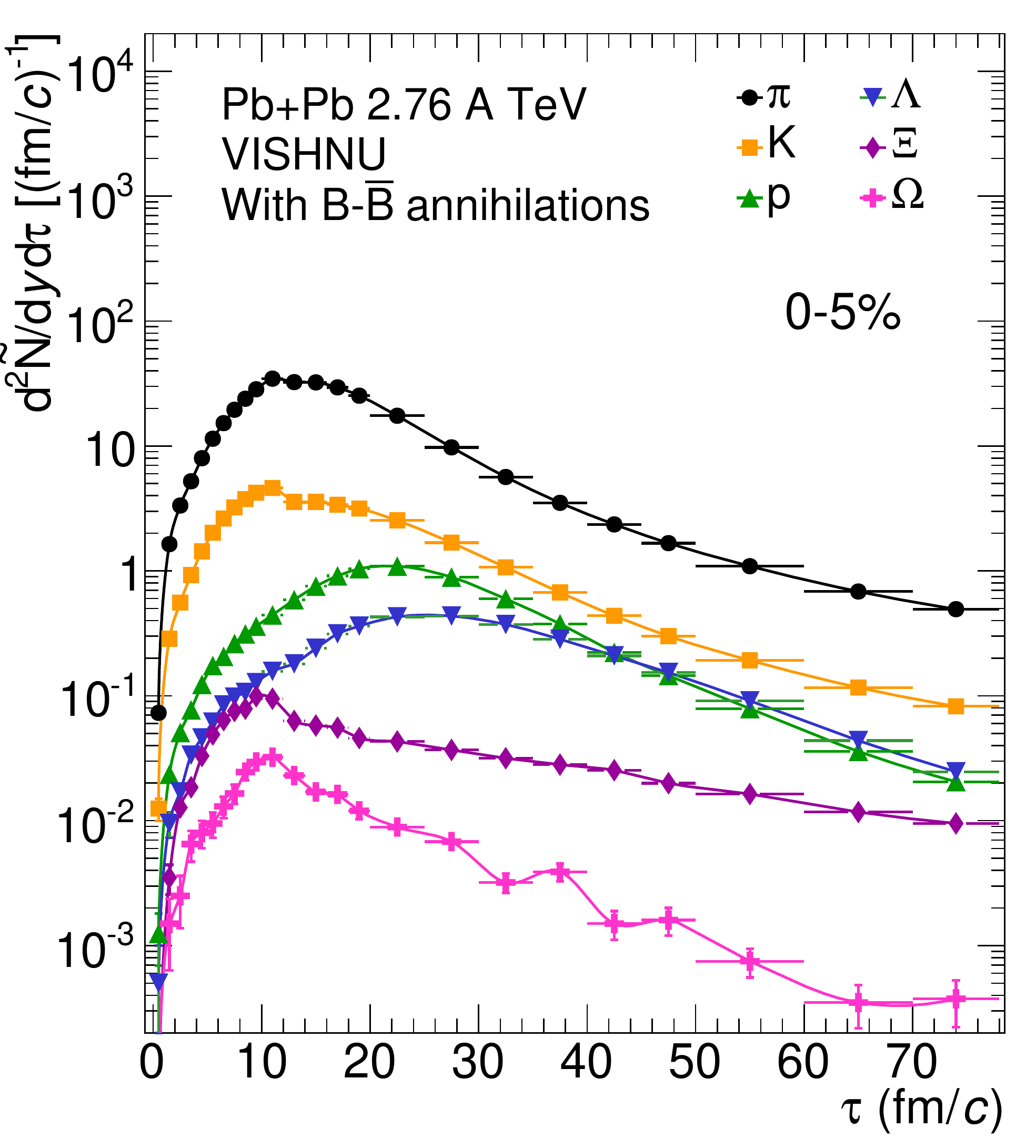}
  \caption{(Color online) Thermal freeze-out time distributions for $\pi$, K, p, $\Lambda$, $\Xi$ and $\Omega$ in the most central Pb+Pb collisions, calculated from {\tt VISHNU} with ${\it B}$-$\bar{B}$ annihilations.
 \label{fig:WBB05}}
\end{figure}

Figure~9 compares the thermal freeze-out time distributions for $\pi$, K, p, $\Lambda$, $\Xi$ and $\Omega$ in the most central Pb+Pb collisions. These curves are calculated from {\tt VISHNU} with ${\it B}$-$\bar{B}$ annihilations, which are the same as the corresponding ones shown in different panels of Fig.~7. The peaks of the $\Xi$ and $\Omega$ curves are both located around 10 ${\rm fm}/c$. Compared with these curves of p and $\Lambda$, which peaks are located around 20-30 ${\rm fm}/c$, these two multi-strange hadrons experience earlier thermal freeze-out. Although the peaks of the $\pi$ and K curves are closer to the ones of $\Xi$ and  $\Omega$, their freeze-out time distributions spread widely along the time axis. This indicates that these two meson species still surfer a certain amount of hadronic scatterings even during the late evolution of {\tt UrQMD}. We conclude from Fig.~9 that thermal freeze-out is hadron species dependent.  Compared with other hadrons, the two multi-strange hadrons $\Xi$ and $\Omega$ experience earlier thermal freeze-out, as expected, due to their much smaller hadronic cross sections.

\section{Summary and outlook\label{sec:summary}}
In this paper, we studied the soft hadron data of strange and multi-strange hadrons at the LHC, using {\tt VISHNU} hybrid model.
We found that, with MC-KLN initial conditions, $\eta/s =0.16$ and other inputs that fit the related data of common hadrons~\cite{Song:2013qma}, {\tt VISHNU} generally describes the multiplicity, $\pt$-spectra and differential elliptic flow of the strange hadron $\Lambda$ and the multi-strange hadrons $\Xi$ and $\Omega$ in 2.76 A TeV Pb+Pb collisions.

Compared with the pure hydrodynamic calculations from {\tt VISH2+1}~\cite{Huovinen:2001cy}, {\tt VISHNU} improves the descriptions of the elliptic flow for p, $\Lambda$, $\Xi$ and $\Omega$ with microscopic hadronic scatterings that re-balance the interactions between radial and elliptic flow. However, mass ordering of elliptic flow among various hadron species is not fully described. {\tt VISHNU} slightly under-predicts the differential elliptic flow of protons, leading to inverse mass-ordering among $p$, $\Lambda$ and $\Xi$. An initial flow and improved hadronic cross sections in {\tt UrQMD} may help to solve this issue. This should be investigated in the near future.

With a nice description of the particle yields for $\pi$, K, p, $\Lambda$, $\Xi$ and $\Omega$, we further investigated chemical and thermal freeze-out of various hadron species within the framework of {\tt VISHNU} hybrid model. We found that, compared with other hadrons, the two multi-strange hadrons, $\Xi$ and $\Omega$ experience earlier chemical and thermal freeze-out due to their small hadronic cross sections. A study for time evolution of the hadron yields also shows that baryon-antibaryon annihilations in {\tt UrQMD} delay the chemical freeze-out of $\Xi$ and $\Omega$, when compared with the case without such annihilations.

We also emphasized that the switching temperature in {\tt VISHNU} could not be identified as the chemical freeze-out temperature in the statistical model since inelastic collisions are still frequent during the early evolution of {\tt UrQMD}. A further analysis of the space-time distributions of the last inelastic collisions could reveal more information on chemical freeze-out, which may even help us to extract effective chemical freeze-out temperatures of various hadron species. Unfortunately, such investigation could not be done with the current version of {\tt UrQMD}. With an updating of {\tt UrQMD} to further record the intermediate evolution information, the chemical freeze-out procedure of the evolving hadronic system will be further studied in the future.

\acknowledgments
\vspace*{-2mm}
We thanks the discussions from S. A. Bass, P. Huovinen, C. Shen and N. Yu. This work was supported by the NSFC and the MOST under grant Nos. 11435001 and 2015CB856900. We gratefully acknowledge extensive computing resources provided to us on Tianhe-1A by the National Supercomputing Center in Tianjin, China.

\end{document}